\documentclass[journal]{IEEEtran}
\usepackage{amsmath,amsfonts,amssymb}
\usepackage{algorithm}
 \usepackage{algpseudocode}
\usepackage{array}
\usepackage{textcomp}
\usepackage{stfloats}
\usepackage{url}
\usepackage{verbatim}
\usepackage{graphicx}
\usepackage{subfigure}
\usepackage{cite}
\usepackage{multirow}
\usepackage{pifont}
\usepackage{float}
\usepackage[]{colortbl}
\usepackage[table]{xcolor}
\usepackage{subfigure, multirow, url, amsfonts, amssymb, pifont, booktabs, siunitx, makecell, hyperref}
\newcommand{\rev}[1]{#1}
\newcommand{\rrev}[1]{#1}

\begin{document}

\title{UniPASE: A Generative Model for Universal Speech Enhancement with High Fidelity and Low Hallucinations}
\author{Xiaobin Rong, Zheng Wang, Yushi Wang, Jun Gao, Jing Lu,~\IEEEmembership{Senior Member,~IEEE}

\thanks{This work was supported by the National Natural Science Foundation of China (Grant No. 12274221), Yangtze River Delta Science and Technology Innovation Community Joint Research Project (Grant No. 2024CSJGG1100), and the AI \& AI for Science Project of Nanjing University. (Corresponding author: Jing Lu.)

The authors are with the Key Laboratory of Modern Acoustics, Institute of Acoustics, Nanjing University, Nanjing 210093, China, and also with the NJU-Horizon Intelligent Audio Lab, Horizon Robotics, Beijing 100094, China (e-mail: xiaobin.rong@smail.nju.edu.cn; zheng.wang@smail.nju.edu.cn; yushi.wang@smail.nju.edu.cn; jun.gao@smail.nju.edu.cn; lujing@nju.edu.cn).}}

\maketitle

\begin{abstract}
Universal speech enhancement (USE) aims to restore speech signals from diverse distortions across multiple sampling rates. We propose UniPASE, an extension of the low-hallucination PASE framework tailored for USE. At its core is DeWavLM-Omni, a unified representation-level enhancement module fine-tuned from WavLM via knowledge distillation on a large-scale supervised multi-distortion dataset. This module directly converts degraded waveforms into clean and linguistically faithful phonetic representations, ensuring robust enhancement with minimal linguistic hallucination. Based on these enhanced phonetic representations, an Adapter generates \rev{enhanced} acoustic representations containing rich acoustic details, which a neural Vocoder uses to reconstruct corresponding high-fidelity 16-kHz waveforms. A PostNet then \rev{converts the waveforms to 48~kHz} before resampling them to their original rates, enabling seamless handling of inputs and outputs at multiple sampling rates.
Experimental results on several evaluation datasets, covering sub-tasks and full tasks, demonstrate that UniPASE achieves superior or competitive performance compared with existing state-of-the-art (SOTA) models. The proposed model also serves as the backbone of our submission to the URGENT 2026 Challenge, which achieved 1st place in the objective evaluation. \rev{The source code and} audio demos are available at \rev{\url{https://github.com/xiaobin-rong/unipase/}}.

\end{abstract}


\section{Introduction}
\IEEEPARstart{U}{niversal} speech enhancement (USE) aims to restore speech signals degraded by various distortions and presented in different input formats (e.g., multiple sampling rates and varying numbers of channels)~\cite{USE}. Recently, the Universality, Robustness, and Generalizability of speech EnhancemeNT (URGENT) Challenge~\cite{URGENT2024, URGENT2025} was launched to advance the development of USE systems and to establish a benchmark for their evaluation. In its second edition, URGENT 2025, the challenge defined a concrete task: to build a single model capable of handling seven types of distortions while supporting flexible sampling rates. The distortions include additive noise, reverberation, clipping, bandwidth limitation, codec artifacts, packet loss, and wind noise. The supported sampling rates cover 8, 16, 22.05, 24, 32, 44.1, and 48 kHz. 

The top-ranked systems in the URGENT 2025 Challenge achieved strong overall performance, yet they were exclusively built upon predictive architectures~\cite{URGENT2025_rank1_tencent} or hybrid predictive–generative designs~\cite{URGENT2025_rank2_nju, URGENT2025_rank3_ut, URGENT2025_rank5_ntu, URGENT2025_rank6_bytedance}. In contrast, purely generative approaches, while capable of delivering superior perceptual quality, remain fundamentally constrained by their vulnerability to \emph{hallucinations}—producing incorrect spoken content or inconsistent speaker characteristics, referred to as \emph{linguistic} and \emph{acoustic} hallucinations, respectively~\cite{PASE}. 
This hallucination issue has emerged as a critical bottleneck that limits the practicality and reliability of purely generative systems, as it violates the core requirement of authenticity. Nevertheless, existing generative approaches often prioritize perceptual quality while underestimating the severity of hallucination effects. For example, a leading purely generative system in the URGENT 2025 challenge achieved the highest perceptual scores but suffered from substantial declines in speaker similarity and character accuracy~\cite{URGENT2025}, highlighting its limited reliability.

Our earlier work, Phonologically Anchored Speech Enhancer (PASE)~\cite{PASE}, was specifically designed to address this issue, introducing a low-hallucination generative speech enhancement (SE) paradigm. To suppress linguistic hallucinations, it leverages the phonological prior encoded in the self-supervised model WavLM~\cite{WavLM} to guide denoising in the \emph{phonetic representation} domain. To mitigate acoustic hallucinations, PASE adopts a dual-stream reconstruction strategy, in which waveform synthesis is primarily driven by the enhanced phonetic representations and explicitly conditioned on low-level noisy \emph{acoustic representations}, thereby preserving speaker characteristics during generation.

Building on this progress, we propose \textbf{UniPASE}, a \textbf{Uni}fied framework that extends \textbf{PASE} to the USE setting established by the URGENT 2025 Challenge. 
First, to address multi-distortion scenarios, we leverage the phonological prior to perform \rev{USE} in the phonetic representation domain, \rev{producing enhanced phonetic representations that are trained to approximate clean, distortion-free targets. This is} motivated by the \rev{idea} that this powerful prior can handle not only noise and reverberation but may generalize to a broader range of distortions. Second, \rev{since the acoustic representations are not explicitly optimized for enhancement in the previous stage,} we introduce an explicit acoustic enhancement stage \rev{that restores low-level acoustic features suitable for waveform synthesis.} \rev{In this stage}, the enhanced phonetic representations \rev{are conditioned on the degraded acoustic representations to generate enhanced} acoustic representations, \rev{which are} then converted into the waveform via a neural vocoder.
Finally, to accommodate variable sampling rates, we incorporate a post-processing module that extends the vocoder output from 16~kHz to 48~kHz and subsequently downsamples it to the desired rate.
To summarize, our contributions are threefold:
\begin{itemize}
    \item We propose UniPASE \rrev{as a system-level extension of} the low-hallucination PASE framework for universal speech enhancement. \rrev{It enables} full-stack speech restoration with support for variable sampling rates, achieving low hallucination and high fidelity.
    \item We introduce an explicit acoustic enhancement stage to complement phonetic enhancement. This stage maps degraded acoustic representations to \rev{enhanced} ones that preserve rich acoustic details, thereby improving speaker fidelity and perceptual quality.
    \item UniPASE \rrev{achieves superior or competitive performance across multiple evaluation datasets compared with advanced baselines}, highlighting the effectiveness of pure generative methods, and demonstrating strong robustness and generalization across different tasks and languages.
\end{itemize}

\section{Related Work}

\begin{table*}[t]
\centering
\caption{Summary and comparison of representative universal speech enhancement systems. Prior systems operate at a single sampling rate, whereas recent URGENT systems are sampling-frequency-independent (SFI).}
\label{tab:use_summary}
\resizebox{\linewidth}{!}{
\begin{tabular}{@{}ccccccccccc@{}}
\toprule
\multirow{2}{*}{Model} & \multirow{2}{*}{Year} & \multicolumn{7}{c}{Distortions} & \multirow{2}{*}{\begin{tabular}[c]{@{}c@{}}Sampling \\ rate (Hz)\end{tabular}} & \multirow{2}{*}{\begin{tabular}[c]{@{}c@{}}Open-sourced?\end{tabular}} \\ \cmidrule(lr){3-9}
 &  & \multicolumn{1}{l}{Additive noise} & \multicolumn{1}{l}{Reverb} & \multicolumn{1}{l}{Clipping} & \multicolumn{1}{l}{Bandwidth limitation} & \multicolumn{1}{l}{Codec artifact} & \multicolumn{1}{l}{Packet loss} & \multicolumn{1}{l}{Wind noise} &  &  \\ \midrule
VoiceFixer~\cite{Voicefixer} & 2022 & \ding{51} & \ding{51} & \ding{51} & \ding{51} &  &  &  & 44.1k & \ding{51} \\
UNIVERSE~\cite{UNIVERSE} & 2023 & \ding{51} & \ding{51} & \ding{51} & \ding{51} & \ding{51} & \ding{51} &  & 16k & \ding{55} \\ 
UNIVERSE++~\cite{UNIVERSE++} & 2024 & \ding{51} & \ding{51} & \ding{51} & \ding{51} & \ding{51} & \ding{51} &  & 24k & \ding{51} \\ 
MaskSR~\cite{MaskSR} & 2024 & \ding{51} & \ding{51} & \ding{51} & \ding{51} & &  &  & 44.1k & \ding{55} \\ 
AnyEnhance~\cite{AnyEnhance} & 2025 & \ding{51} & \ding{51} & \ding{51} & \ding{51} & &  &  & 44.1k & \ding{51} \\ 
LLaSE-G1~\cite{LLaSE-G1} & 2025 & \ding{51} & \ding{51} &  &  & & \ding{51} &  & 16k & \ding{51} \\ \midrule
SIG systems~\cite{SIG2_KSNet, SIG2_RADNet, SIG2_NJUNet} & 2024 & \ding{51} & \ding{51} & \ding{51} & \ding{51} & \ding{51} & \ding{51} &  & 24k & \ding{55} \\ \midrule
URGENT systems~\cite{URGENT2025_rank1_tencent, URGENT2025_rank2_nju, URGENT2025_rank3_ut, URGENT2025_rank5_ntu, URGENT2025_rank6_bytedance} & 2025 & \ding{51} & \ding{51} & \ding{51} & \ding{51} & \ding{51} & \ding{51} & \ding{51} & SFI & \ding{55} \\ \midrule
\rev{UniPASE (proposed)} & \rev{2026} & \rev{\ding{51}} & \rev{\ding{51}} & \rev{\ding{51}} & \rev{\ding{51}} & \rev{\ding{51}} & \rev{\ding{51}} & \rev{\ding{51}} & \rev{SFI} & \rev{\ding{51}} \\ \bottomrule
\end{tabular}
}
\end{table*}

\subsection{Universal Speech Enhancement Models}
There has been growing interest in USE frameworks capable of addressing multiple types of speech distortions. In some prior work, similar settings are referred to as general speech restoration (GSR); however, GSR typically does not consider multi-rate processing. A summary and comparison of representative USE systems is provided in Table~\ref{tab:use_summary}. 
VoiceFixer~\cite{Voicefixer}, a pioneering unified framework for speech restoration, comprises a ResUNet-based Mel-domain restoration module and a neural vocoder for waveform synthesis, targeting four distortion types: noise, reverberation, clipping, and bandwidth limitation.
UNIVERSE~\cite{UNIVERSE}, along with its advanced variant UNIVERSE++~\cite{UNIVERSE++}, leverages score-based diffusion to handle a wider range of distortions, additionally covering codec artifact and packet loss.
MaskSR~\cite{MaskSR}, like VoiceFixer, adopts a similar two-stage architecture and targets the same four distortion types, but replaces the restoration module with a more powerful token-based masked generative model (MGM) and substitutes the vocoder with a pre-trained codec, DAC~\cite{DAC}.
AnyEnhance~\cite{AnyEnhance}, which also adopts an MGM, further refines the restoration process by decomposing it into two stages: semantic enhancement and acoustic enhancement, and additionally extends its applicability to target speaker extraction (TSE).
LLaSE-G1~\cite{LLaSE-G1} employs a simpler language modeling strategy, parallel token prediction, while supporting a distinct set of sub-tasks including denoising, dereverberation, packet loss concealment (PLC), TSE, acoustic echo cancellation (AEC), and speech separation (SS).

Despite strong performance across multiple test sets, purely generative models often struggle to top benchmark evaluations. Recent SE challenges demonstrate that predictive or hybrid predictive-generative approaches often achieve higher overall rankings. For instance, in the ICASSP 2024 Speech Signal Improvement (SIG) Challenge, the top-ranked systems~\cite{SIG2_KSNet, SIG2_RADNet, SIG2_NJUNet} all adopt a two-stage “restoration-enhancement” framework, optimized with a combination of predictive and adversarial objectives. This design allows them to exploit the denoising and dereverberation strengths of predictive models while leveraging generative losses to reconstruct missing components, such as those caused by bandwidth limitations or packet loss.

The URGENT 2025 Challenge further extends distortion coverage to include wind noise—a common real-world artifact—and supports multiple input sampling rates. It also introduces an extensive set of evaluation metrics to enable comprehensive assessment. Under this protocol, a purely predictive large-scale BSRNN~\cite{URGENT2025_rank1_tencent} ranked first, excelling on most objective metrics (e.g., intrusive scores, speaker similarity, and character accuracy) despite being suboptimal in perceptual quality (e.g., non-intrusive metrics and subjective MOS). In contrast, the hybrid predictive-generative models TS-URGENet~\cite{URGENT2025_rank2_nju} and FUSE~\cite{URGENT2025_rank3_ut} achieved better perceptual quality but slightly lagged behind the first-ranked system on most objective metrics, ultimately ranking 2nd and 3rd. Consistent with this trend, a purely generative model attained the highest perceptual quality but suffered from severe hallucinations and poor performance on other objective metrics, ultimately placing only 13th~\cite{URGENT2025}. These observations highlight the need for generative models that combine high perceptual quality with low hallucinations.

\begin{figure*}[t]
    \centering
    \includegraphics[width=0.9\linewidth]{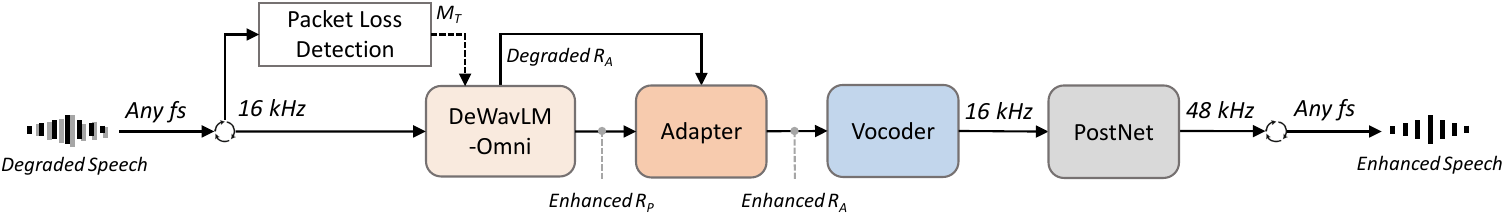}
    \caption{Overview of the proposed UniPASE framework for universal speech enhancement.}
    \label{fig:unipase}
\end{figure*}

\subsection{Speech Enhancement with Self-supervised Models}
\rev{Self-supervised learning (SSL) has demonstrated substantial value for SE. Early predictive approaches leveraged SSL representations to replace or augment spectrogram inputs while still estimating spectrogram masks~\cite{Investigating_SSL_for_SE, Boosting_SSL_SE}. More recent generative methods directly perform core enhancement in the SSL domain, which can be broadly categorized into two paradigms: single-stage and two-stage.}

Single-stage methods enhance an intermediate SSL representation and then use a vocoder for waveform reconstruction~\cite{Miipher, Efficient_SSL_SE, Miipher-2}. The intermediate layer is typically chosen because it simultaneously contains high-level semantic information for linguistically faithful enhancement and fine-grained acoustic details for waveform reconstruction. For example, Miipher~\cite{Miipher} uses a DF-Conformer-based~\cite{DF-Conformer} feature cleaner on the 8th layer of w2v-BERT~\cite{w2v-BERT} representations, while Sun~et al.~\cite{Efficient_SSL_SE} systematically explore different SSL models and feature cleaner architectures. \rrev{In addition, some approaches apply discrete language modeling (LM) for feature cleaning by converting noisy continuous representations into discrete tokens and generating their clean counterparts either in parallel \cite{SELM, LLaSE-G1} or autoregressively \cite{Genhancer}}. 

Two-stage methods decompose enhancement into semantic and acoustic stages~\cite{GenSE, AnyEnhance, SenSE}. \rrev{Within this paradigm, both stages can be realized using various strategies such as LMs~\cite{GenSE, SenSE} and diffusion models~\cite{SISE}}.
These stages operate on semantic representations\footnote{\rev{Also referred to as \emph{phonetic} representations, as they exhibit properties closer to phonetic-like units.~\cite{SSL_phonetic, Comparative_layerwise_SSL_analysis}}} and acoustic representations, with the former typically derived from deep-layer SSL representations, while the latter can take various forms (e.g., codec tokens~\cite{GenSE} and Mel spectrograms~\cite{SenSE}) as long as they preserve sufficient fine-grained acoustic information for waveform reconstruction. The semantic stage generates conditions to guide the subsequent acoustic enhancement, often yielding superior performance over single-stage methods.

\rrev{However, the above approaches all follow a paradigm of cleaning noisy SSL representations, which may be inherently limited since these representations already deviate from desired semantic and acoustic properties due to corruption. In contrast, directly learning noise-invariant SSL representations can better preserve the intended phonetic and acoustic attributes, providing a more robust alternative. For instance,} Miipher-2~\cite{Miipher-2}, a single-stage method, introduces lightweight adapters directly within the SSL backbone USM~\cite{USM}, finetuning only these adapters to clean the 13th layer representations efficiently. \rrev{Our previous work, PASE \cite{PASE}, follows a two-stage design. It first finetunes the whole SSL model WavLM~\cite{WavLM} via distillation to generate clean 24th-layer representations (referred to as phonetic representations), which are then conditioned on 1st-layer representations (referred to as acoustic representations) for waveform reconstruction through a vocoder. This design demonstrates strong semantic integrity, but may suffer from limited acoustic fidelity due to the direct use of noisy acoustic representations in waveform generation.}

\rrev{Motivated by this observation,} in the present work, we first extend PASE's semantic stage to address various degradations beyond noise and reverberation, and further decompose the acoustic stage by enhancing the acoustic representations before generating waveforms with a pretrained neural vocoder, yielding a robust and high-fidelity USE system.

\section{Background: PASE}
The PASE framework consists of a denoising WavLM (DeWavLM) and a vocoder, \rev{corresponding to semantic and acoustic enhancement, respectively}. DeWavLM is created by fine-tuning a pre-trained WavLM through a denoising representation distillation (DRD) strategy, adapting it into a denoising expert. Specifically, we instantiate two copies of WavLM: a frozen teacher and a trainable student, both initialized from the pre-trained weights to inherit the phonological prior. The student model is trained to map a noisy input waveform to a clean representation by minimizing the mean-squared error (MSE) loss against the target representation, which is generated by the teacher from the corresponding clean waveform. \rev{The loss is computed on} the outputs of the final Transformer layer, referred to as \emph{phonetic representations} \rev{(denoted as $\text{R}_{\text{P}}$ hereafter)}, which encode abstract and context-dependent phonetic information:
\begin{equation}
    \rev{\mathcal{L}_{\mathrm{MSE}} =
    \left\|
    R_{P}^{\mathrm{student}}(x_{\mathrm{noisy}})
    -
    R_{P}^{\mathrm{teacher}}(x_{\mathrm{clean}})
    \right\|_2^2}
\label{eq:loss}
\end{equation}

The vocoder reconstructs the enhanced waveform from DeWavLM's dual-stream representations: the phonetic representations and \emph{acoustic representations} \rev{(denoted as $\text{R}_{\text{A}}$ hereafter)}, which are taken from the first Transformer layer and retain fine-grained acoustic details essential for preserving speaker identity and prosody. This design ensures that the synthesized speech preserves both the content and the speaker’s characteristics. \rev{To clarify, the terms ``phonetic" and ``acoustic" refer to the dominant information identified in existing analysis~\cite{Comparative_layerwise_SSL_analysis}, rather than implying strictly disentangled representations.}

\section{UniPASE}
In this section, we present UniPASE, a generative model for USE that achieves high fidelity while mitigating hallucinations. As illustrated in Fig.~\ref{fig:unipase}, UniPASE consists of four key components: DeWavLM-Omni, an Adapter, a Vocoder, and a PostNet. \rev{Compared with PASE, UniPASE retains the same dual-stream generative framework, while replacing DeWavLM with DeWavLM-Omni to support broader distortions, introducing an Adapter for improved perceptual quality, and employing a PostNet to enable flexible sampling rates.}

Given a degraded signal as input, which may be sampled at any rate, the waveform is first resampled to 16~kHz. A packet loss detection \rev{(PLD)} algorithm is then applied to identify missing frames, producing a binary mask $\text{M}_\text{T}$, which is used by DeWavLM-Omni to perform core enhancement in the phonetic representation domain. It takes the degraded waveform along with the packet-loss mask as inputs to conduct universal speech enhancement, producing dual-stream outputs: an enhanced phonetic representation \rev{(denoted as \emph{Enhanced~$\text{R}_{\text{P}}$} in Fig.~\ref{fig:unipase})} \rev{with degradations effectively removed}, and a degraded acoustic representation \rev{(denoted as \emph{Degraded~$\text{R}_{\text{A}}$})}, \rev{which is not explicitly optimized for enhancement}. The Adapter then performs explicit acoustic enhancement, mapping these representations to \rev{an enhanced} acoustic representation \rev{(\emph{denoted as Enhanced~$\text{R}_{\text{A}}$})}, which is subsequently used by the Vocoder to synthesize an enhanced waveform at 16~kHz. Finally, the PostNet \rev{converts the waveform to 48~kHz}, which is then downsampled to match the original rate, \rev{enabling flexible sampling-rate I/O}. Note that the PostNet is only applied when the original rate exceeds 16~kHz. Detailed descriptions of each module are in the following subsections.

\subsection{DeWavLM-Omni}

DeWavLM-Omni extends DeWavLM to the USE setting by adopting the same DRD strategy, while further augmenting the noisy input speech with diverse distortions to encourage the learning of \rev{degradation}-invariant representations \rev{(i.e., learning representations corresponding to clean speech without any degradations)}. This design is motivated by the \rev{idea} that the phonological prior within WavLM can not only guide low-hallucination denoising and dereverberation, but may also generalize to other types of distortions. In particular, since this prior arises from masked-prediction-based pre-training, which enables the model to infer missing regions from contextual information, it is naturally well-suited for guiding PLC.

To explicitly exploit this capability, we employ the packet loss detection (PLD) algorithm to identify missing frames and replace the corresponding CNN output frames in WavLM with \rev{a shared learnable mask embedding}. The PLD algorithm segments the input waveform into short, non-overlapping packets and identifies nearly silent ones. For each packet, we compute the fraction of samples with amplitudes below a small threshold; if this fraction exceeds a predefined ratio, the packet is flagged as lost. The complete algorithm is provided in Appendix~\ref{app:alg}. \rev{It is worth noting that although the PLD algorithm is relatively simple and may occasionally over-detect packet losses (e.g., misclassifying silent segments as lost), DeWavLM-Omni remains robust to such errors. Since the detected frames are replaced with the shared mask embedding, such misdetections introduce only mild perturbations rather than structured distortions. As DeWavLM-Omni is trained to remove various degradations and is exposed to similar patterns (e.g., zero-padded or partially missing speech) during training, it can effectively handle these cases.}

\rev{DeWavLM-Omni follows the loss formulation in Eq.~(\ref{eq:loss}). Notably, the loss is computed over all frames rather than only the masked ones. With this objective, it} produces dual-stream representations, \rev{both with shape $\mathbb{R}^{T\times D}$, where $T$ denotes the number of time frames and $D$ the feature dimensionality}: (1) \rev{degraded~$\text{R}_{\text{A}}$}, from the first Transformer layer \rev{and not explicitly optimized with an MSE loss for enhancement}; and (2) \rev{enhanced~$\text{R}_{\text{P}}$}, from the final Transformer layer, which contains rich and purified phonetic information.

\subsection{Adapter}

Although directly reconstructing the waveform from the dual-stream representations using a neural vocoder is effective in PASE, it can result in noise and reverberation leakage when the input signal-to-noise ratio (SNR) is low. To mitigate this, we introduce an explicit acoustic enhancement stage implemented by an Adapter, which restores low-level acoustic details before waveform synthesis, thereby improving the final reconstruction quality. \rev{Specifically, the Adapter takes the degraded $\text{R}_{\text{A}}$ as input and conditions on the enhanced representation $\text{R}_{\text{P}}$ to produce the enhanced~$\text{R}_{\text{A}}$. Following ablation findings in PASE, the conditioning is implemented via element-wise summation, which is simple yet effective. The training target is the clean $\text{R}_{\text{A}}$ from DeWavLM-Omni given clean speech, without any degradations.}

The Adapter is based on the improved Vocos~\cite{Vocos} backbone proposed in~\cite{WavTokenizer}, which integrates an attention module to enhance contextual modeling. Since it is tasked with generating fine-grained acoustic details from highly abstract phonetic representations, relying solely on a standard regression loss (e.g., MSE) can cause over-smoothing and yield representations with diminished structural detail. To mitigate this issue, we introduce an adversarial objective with a representation-domain discriminator, termed Multi-\rrev{Resolution} Representation Discriminator (\rrev{MRRD}), which is designed to \rev{model the representations} across multiple \rev{feature} \rrev{resolutions}.

\begin{figure}
    \centering
    \includegraphics[width=0.9\linewidth]{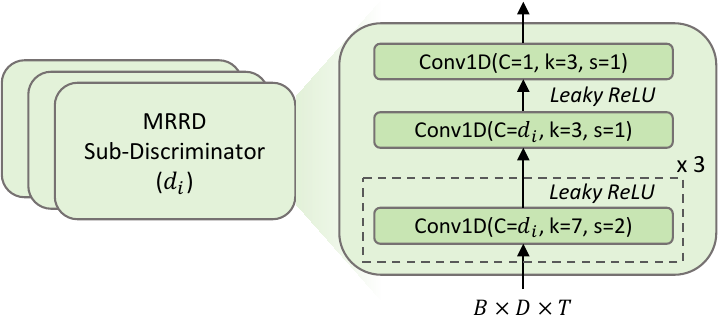}
    \caption{Multi-\rrev{Resolution} Representation Discriminator (\rrev{MRRD}).}
    \label{fig:disc}
\end{figure}

As illustrated in Fig.~\ref{fig:disc}, the \rrev{MRRD} is composed of multiple sub-discriminators, each responsible for \rev{modeling the input representations at a specific \emph{feature \rrev{resolution}} (i.e., the hidden dimensionality of the processed representations)}. Each sub-discriminator is implemented as a stack of 1D convolutional layers with \rev{Leaky ReLU}~\cite{LeakyReLU}, where the first layer projects the input to a hidden space of $d_i$ channels, \rev{corresponding to its feature \rrev{resolution}. Intuitively, sub-discriminators with smaller hidden dimensions may capture more global or coarse characteristics of the representations, while those with larger dimensions may be sensitive to finer details.} By aggregating feedback across \rev{different feature} \rrev{resolutions}, \rrev{MRRD} \rev{provides guidance to} the generator \rev{for producing more} structurally faithful outputs.

The training objective follows LS-GAN~\cite{LS-GAN}, which alleviates the vanishing gradient issue. The adversarial losses for the generator $G$ and the discriminator $D$ are defined as:
\begin{align}
    \mathcal{L}_{adv}(G;D) &= \mathbb{E}_{\hat{r}} \Big[ (D(\hat{r}) - 1)^2 \Big], \\
    \mathcal{L}_{adv}(D;G) &= \mathbb{E}_{(r,\hat{r})} \Big[ (D(r)-1)^2 + (D(\hat{r}))^2 \Big],
\end{align}
where $r$ and $\hat{r}$ denote the ground truth and the generated representations, respectively, \rev{each of shape $\mathbb{R}^{T\times D}$}.
Following~\cite{HiFiGAN}, additional reconstruction and feature-matching terms are incorporated in the training of the generator. 
The reconstruction loss $\mathcal{L}_{rec}$ is implemented as MSE:
\begin{equation}
    \mathcal{L}_{rec}(G) = \mathbb{E}_{(r,\hat{r})} \Big[ (r - \hat{r})^2 \Big],
\end{equation}
and the feature-matching loss $\mathcal{L}_{feat}$ is computed as the L1 distance between the $l$th feature maps of the $k$th sub-discriminator from the ground-truth and the generated samples:
\begin{equation}
    \mathcal{L}_{feat}(G;D) =\mathbb{E}_{(r,\hat{r})}\Big[ \textstyle \frac{1}{KL} \sum_{k} \sum_{l} \left|D_k^l(r) - D_k^l(\hat{r}) \right| \Big].
\end{equation}
The final objectives for the generator and discriminator are:
\begin{align}
    \mathcal{L}_G &= \mathcal{L}_{adv}(G;D) + \lambda_{feat} \mathcal{L}_{feat}(G;D) + \lambda_{rec} \mathcal{L}_{rec}(G), \\
    \mathcal{L}_D &= \mathcal{L}_{adv}(D;G),
\end{align}
where $\lambda_{feat}$ and $\lambda_{rec}$ are set to 1 and 200, respectively.

\subsection{Vocoder}
\rev{Since the acoustic representations are limited to 16~kHz, the Vocoder is designed to reconstruct the enhanced waveform at a 16~kHz sampling rate.} It is trained independently on clean speech and subsequently integrated into the system without any joint fine-tuning. The Vocoder also adopts the improved Vocos\rev{~\cite{WavTokenizer}} architecture \rev{to process the acoustic representations $\text{R}_{\text{A}}\in\mathbb{R}^{T\times D}$}, and incorporates an iSTFT head for waveform synthesis. Its training objective follows that of the vocoder in PASE, consisting of a reconstruction loss implemented via multi-scale Mel-spectrogram loss~\cite{DAC}, combined with adversarial and feature-matching losses computed using a multi-period discriminator (MPD) and a multi-band multi-scale STFT discriminator (MBMSD)~\cite{DAC}. 

\subsection{PostNet}
As the final post-processing module of UniPASE, the PostNet \rev{is used to upsample the 16-kHz Vocoder output when the desired sampling rate exceeds 16~kHz. Specifically, it first generates a 48-kHz waveform through bandwidth extention (BWE), which is then downsampled to match the original rate.} It follows the \rev{STFT-domain} CWS-TF-GridNet architecture from TS-URGENet~\cite{URGENT2025_rank2_nju}, which combines the channel-wise subband (CWS)~\cite{CWS} and TF-GridNet, and is optimized with the same loss used for the Vocoder, including multi-scale Mel-spectrogram, adversarial, and feature-matching terms.

While residual connections are commonly used in BWE to preserve low-frequency components, they can still alter the low-band spectrum and degrade perceptual quality. To address this, we explicitly retain the low-frequency components by directly copying them from the input spectrogram during inference, thereby minimizing the impact of BWE on the reliable low-band content. To ensure smooth spectral integration at the band boundary, a transition band is applied between the copied low-frequency region and the reconstructed high-frequency band. Given an input spectrogram $X(f,t)\in \mathbb{C}$ \rev{derived from the 16-kHz Vocoder output and resampled to 48~kHz}, the network outputs a full-band spectrogram $H(f,t) \in \mathbb{C}$, where $f$ and $t$ denote the frequency bins and time frames. The final output full-band spectrogram $Y(f,t) \in \mathbb{C}$ is computed as
\begin{equation}
Y(f,t) = X(f,t) + \alpha(f)\, H(f,t),
\end{equation}
with
\begin{equation}
\alpha(f) =
\begin{cases}
0, & f \le f_c - \Delta f,\\
(f-f_c + \Delta f)/\Delta f, & f_c - \Delta f < f \le f_c,\\
1, & f > f_c,
\end{cases}
\end{equation}
\rev{Here, $f_c = 8$~kHz is fixed by the intrinsic bandwidth limitation of the preceding modules (i.e., the DeWavLM-Omni backbone and Vocoder), whose output has an effective bandwidth of 8~kHz regardless of the input bandwidth or corruption type. The transition bandwidth $\Delta f$ is set empirically to $800$~Hz to minimize low-frequency alteration and spectral discontinuities.}

\subsection{\rev{Overall Training Procedure}}
\rev{UniPASE is trained in a staged manner. The Vocoder is trained independently on clean speech. Separately, DeWavLM-Omni is trained on degraded speech, with packet-loss indicators provided by the PLD algorithm. The Adapter is then trained on top of DeWavLM-Omni while keeping the latter frozen. Finally, the PostNet is trained on top of the preceding cascaded pipeline (DeWavLM-Omni, Adapter, and Vocoder), with all these modules kept frozen.}

\section{Experiments}
\subsection{Datasets}
\subsubsection{Training Datasets}

The training dataset is constructed from large-scale corpora provided by the URGENT 2025 Challenge~\cite{URGENT2025}. Clean speech is collected from multiple sources, including the LibriVox subset of the DNS5 Challenge~\cite{DNS5}, LibriTTS~\cite{LibriTTS}, VCTK~\cite{VCTK}, EARS~\cite{EARS}, MLS~\cite{MLS}, and Common Voice 19.0~\cite{CommonVoice}.
To ensure high-quality training data, all corpora except EARS are filtered using DNSMOS scores (OVRL, SIG, BAK, and P.808) with a threshold of 3.0. EARS is excluded from this filtering process because DNSMOS is unreliable for atypical speech, such as whispers or speech with extreme pitch. Manual inspection further confirms that the EARS recordings are consistently of high quality. After filtering, the final clean speech dataset contains approximately 2,360 hours of audio.
Noise samples are drawn from DNS5, WHAM!~\cite{WHAM}, FSD50K~\cite{FSD50K}, FMA~\cite{FMA}, as well as a simulated wind noise database. 
Room impulse responses (RIRs) are taken from openSLR26 and openSLR28~\cite{openSLR}. 

Training mixtures are generated on the fly using the official data preparation scripts\footnote{\url{https://github.com/urgent-challenge/urgent2025_challenge/tree/main/simulation}}. 
For each \rev{sample, the target is the original clean utterance, and the noisy input is obtained by convolving} the clean utterance with a randomly selected RIR with probability 0.5\rev{, followed by mixing it} with a randomly selected noise clip at an SNR uniformly sampled between -5 and 15~dB.
With probability 0.05, the noise is drawn from a simulated wind noise database and \rev{added using the non-linear mixing simulator provided in~\cite{WindNoise}}; otherwise, it is sampled from the remaining noise sources and added directly.

The resulting mixture is further subjected to distortion augmentations. Specifically, zero, one, two, or three augmentations are applied with probabilities of 0.25, 0.40, 0.20, and 0.15, respectively. Four types of distortions (clipping, bandwidth limitation, codec artifact, and packet loss) are selected with equal probability. Detailed hyperparameter settings for all augmentations are summarized in Table~\ref{tab:aug}.

\begin{table}[t]
\centering
\caption{Data augmentation configurations. $^\dagger$These probabilities are derived as the expected values under the multi-augmentation sampling scheme.}
\begin{tabular}{@{}lcc@{}}
\toprule
Distortion & Probability & Hyperparameters \\ \midrule
Reverberation & 0.5 & - \\ \midrule
Noise & 0.95 & SNR $\in$ [-5, 15]~dB \\ \midrule
Wind noise & 0.05 & SNR $\in$ [-5, 15]~dB \\ \midrule
Clipping & 0.3125$^\dagger$ & \begin{tabular}[c]{@{}c@{}}Min\_quantile $\in$ [0.0, 0.1]\\ Max\_quantile $\in$ [0.9, 1.0]\end{tabular} \\ \midrule
Bandwidth limitation & 0.3125$^\dagger$ & Bandwidth = 4 kHz \\ \midrule
Codec artifact & 0.3125$^\dagger$ & \begin{tabular}[c]{@{}c@{}}Format $\in$ \{mp3, ogg\}\\ Qscale $\in$ [-1, 10]\end{tabular} \\ \midrule
Packet loss & 0.3125$^\dagger$ & \begin{tabular}[c]{@{}c@{}}Duration = 20 ms\\ Rate $\in$ [0.05, 0.25]\\ Max\_continuous\_loss = 10\end{tabular} \\  \bottomrule
\end{tabular}
\label{tab:aug}
\end{table}

\subsubsection{Evaluation Datasets}
We employ multiple evaluation datasets to comprehensively assess model performance, with each dataset corresponding to a specific task. 
\begin{itemize}
    \item \textbf{DNS 2020 test set}: The official synthetic test set from the Interspeech 2020 DNS Challenge~\cite{DNS1}, used to evaluate conventional speech enhancement tasks, including denoising and dereverberation. The dataset consists of two subsets, \textit{with-reverb} and \textit{no-reverb}, depending on whether the clean speech contains reverberation. Notably, for the with-reverb subset, we use the corresponding no-reverb clean utterances as references when computing evaluation metrics, as our evaluation also targets dereverberation performance. All samples are provided at 16~kHz.

    \item \textbf{PLC 2024 validation set}: The official synthetic validation set from the ICASSP 2024 PLC Challenge~\cite{PLC2024}, used to evaluate PLC performance. It provides paired lossy and clean utterances together with lost-frame annotations; however, these annotations are not used during evaluation. All audio is sampled at 48~kHz.

    \item \textbf{VoiceFixer GSR test set}: The open-source test set from VoiceFixer~\cite{Voicefixer}, used to evaluate a broad range of speech restoration tasks, including denoising, dereverberation, declipping, and BWE. All audio is sampled at 44.1~kHz.

    \item \textbf{URGENT 2025 non-blind test set}: The official non-blind test set from the URGENT 2025 Challenge~\cite{URGENT2025}, used to evaluate USE performance. It includes a wide range of distortions and audio sampled at diverse rates, requiring the model to perform full-stack speech restoration \rev{(i.e., addressing all the distortions listed in Table~\ref{tab:use_summary})} while accommodating varying input sampling rates.
    
\end{itemize}

\subsection{Baselines}
We compare UniPASE against a diverse set of SOTA baselines across different evaluation tasks and datasets:

\begin{itemize}
    \item \textbf{SE baselines}: the predictive TF-GridNet~\cite{TF-GridNet}, the diffusion-based StoRM~\cite{StoRM}, the LM-based LLaSE-G1~\cite{LLaSE-G1}, the MGM-based AnyEnhance~\cite{AnyEnhance}, and our previous work PASE~\cite{PASE}.

    \item \textbf{PLC baselines}: TF-GridNet, LLaSE-G1 and the diffusion-based UNIVERSE++~\cite{UNIVERSE++}.

    \item \textbf{GSR baselines}: TF-GridNet, VoiceFixer~\cite{Voicefixer}, and AnyEnhance.

    \item \textbf{USE baselines}: submitted systems from the URGENT 2025 Challenge, including a large-scale BSRNN augmented with Fourier Analysis Networks (FAN)~\cite{URGENT2025_rank1_tencent} (denoted as BSRNN-FAN in this paper for convenience), TS-URGENet~\cite{URGENT2025_rank2_nju}, FUSE~\cite{URGENT2025_rank3_ut}, USEMamba~\cite{URGENT2025_rank5_ntu}, the official baseline TF-GridNet, and a representative purely generative system \rev{from Team \texttt{wataru9871}, as reported in~\cite{URGENT2025}}.
\end{itemize}

For TF-GridNet, we use the released checkpoint provided as the official baseline in URGENT 2025 Challenge\footnote{\url{https://huggingface.co/kohei0209/tfgridnet_urgent25}}. 
For StoRM, LLaSE-G1, and VoiceFixer, we employ their officially released checkpoints\footnote{\url{https://github.com/sp-uhh/storm}}$^,$\footnote{\url{https://huggingface.co/ASLP-lab/LLaSE-G1}}$^,$\footnote{\url{https://github.com/haoheliu/voicefixer}}.
Notably, StoRM provides separate checkpoints for denoising (\texttt{WSJ+CHiME3}) and dereverberation (\texttt{WSJ+Reverb}). To enable joint evaluation on the DNS 2020 with-reverb test set, we sequentially apply dereverberation followed by denoising during inference. 
For AnyEnhance, metrics are computed using the inference audio provided by the authors. 
For UNIVERSE++, we use the metric results reported in the original paper. 
For USE baselines, we adopt the results from the official URGENT 2025 leaderboard\footnote{\url{https://urgent-challenge.com/competitions/13\#results}}. \rev{It is worth noting that these scores are obtained from the non-blind test leaderboard and reflect model performance during the development phase rather than the final submitted systems. Therefore, these results are provided only for reference.}

\subsection{Evaluation Metrics}
Following the URGENT 2025 Challenge, we report a comprehensive set of evaluation metrics spanning perceptual quality, intelligibility, speaker similarity, and linguistic correctness:
\begin{itemize}
    \item \textbf{Non-intrusive metrics}: DNSMOS~\cite{DNSMOS-P835} (16~kHz), UTMOS~\cite{UTMOS} (16~kHz), and NISQA~\cite{NISQA} (48~kHz). PLCMOS~\cite{PLC2022} is additionally reported on the PLC 2024 validation set.

    \item \textbf{Intrusive metrics}: Perceptual evaluation of speech quality (PESQ)~\cite{PESQ} and extended short-time objective intelligibility (ESTOI)~\cite{ESTOI}. These metrics are sensitive to imperceptible signal-level differences and may be unreliable for generative models~\cite{PESQ_unreliable, LPS}.

    \item \textbf{Representation-similarity-based metrics}: Speaker similarity (SpkSim), Levenshtein phoneme similarity (LPS)~\cite{LPS}, and SpeechBERTScore (SBS)~\cite{SpeechBERTScore}, computed as cosine similarity between pretrained representations of enhanced and reference speech.

    \item \textbf{ASR-based metrics}: Word error rate (WER) or character error rate (CER), depending on whether the dataset is English-only or multilingual. When reference transcripts are unavailable, ASR transcriptions of clean speech are used as pseudo-references, yielding dWER or dCER.
\end{itemize}

Consistent with URGENT 2025, we employ \rev{HuBERT~\cite{HuBERT} for SBS on English data and mHuBERT-147~\cite{mHuBERT-147} on multilingual data}, a wav2vec~2.0 model fine-tuned for phoneme recognition~\cite{Wav2Vec2-LV60K-ft} for LPS, and RawNet3~\cite{RawNet3} for SpkSim.
For ASR-based metrics, OWSM v3.1~\cite{OWSM} is used on the URGENT 2025 evaluation set to ensure fair comparison, while Whisper-Large-v3~\cite{Whisper-Large-v3} is adopted for all other test sets due to its superior robustness.

\subsection{Implementation Details}
This section presents the implementation details of UniPASE, summarized as follows:

\textbf{PLD configuration}: The packet loss detection (PLD) algorithm uses a packet duration of 20~ms, consistent with the data augmentation pipeline. The amplitude threshold is set as 1e-4, and the minimum zero ratio is 0.99.
    
\textbf{DeWavLM-Omni configurations}: The DeWavLM-Omni module adopts the WavLM-Large configuration. During distillation training, all model parameters are updated. 
    
\textbf{Adapter configurations}: The Adapter follows the improved Vocos\rev{~\cite{WavTokenizer}} architecture, comprising a hidden dimension of 1024, 4 ResNet blocks, an attention module, and 12 ConvNeXt~\cite{ConvNeXt} blocks with a shared intermediate dimension of 3072. The \rrev{MRRD} discriminator contains six sub-discriminators with progressively increasing hidden channels of $[32, 64, 128, 256, 512, 1024]$. The loss weights for reconstruction, adversarial, and feature-matching terms are empirically set to 200, 1, and 1, respectively. The Adapter is trained \rev{on top of} DeWavLM-Omni, with DeWavLM-Omni parameters frozen during this stage.
    
\textbf{Vocoder configurations}: The Vocoder adopts the same backbone as the Adapter, augmented with an iSTFT for waveform synthesis (FFT size 1280, hop size 320). For the multi-scale Mel-spectrogram reconstruction loss, we use window lengths of $[32, 64, 128, 256, 512, 1024, 2048]$ and corresponding Mel bins of $[5, 10, 20, 40, 80, 160, 320]$, with hop lengths set to $1/4$ of the window length. The reconstruction, adversarial, and feature-matching losses are weighted 30, 1, and 1, respectively. The Vocoder is trained independently and then integrated into the system without joint fine-tuning.
    
\textbf{PostNet configurations}: The PostNet adopts the CWS-TF-GridNet architecture from TS-URGENet~\cite{URGENT2025_rank2_nju}, with an FFT size of 1536, hop size of 768, embedding dimension of 48, LSTM hidden dimension of 100, 4 attention heads, and a total of 5 blocks. The cutoff frequency $f_c = 8$~kHz corresponds to 256 frequency bins, and the transition bandwidth $\Delta f = 800$~Hz spans 24 bins. The loss weights follow the same configuration as those used in the Vocoder training. As the final module of the whole system, the PostNet is trained trained \rev{on top of} the preceding modules, while keeping all of them frozen.

\textbf{Loss weight principle}: Across all stages, loss weights are empirically set such that the reconstruction term dominates, while adversarial and feature-matching terms contribute approximately $1/10$ of the total loss. This design promotes stable training and preserves high-fidelity outputs.

\textbf{Training setup}: All models are trained on four NVIDIA 4090 GPUs using the AdamW~\cite{AdamW} optimizer, with a linear warm-up \rev{from zero to the peak learning rate} over the first 10\% of steps, followed by cosine decay to 1e-6 over the remaining steps. During training, all utterances are resampled to a target sampling rate and randomly cropped or padded to a fixed length; only utterances with an original sampling rate higher than the target are used. Table~\ref{tab:conf} summarizes additional details, including the number of parameters (Params), computational cost (MACs\footnote{Computed using the \texttt{ptflops} toolkit: \url{https://pypi.org/project/ptflops}} per second), sampling rate (Fs), utterance length (Len), batch size per GPU (Bs), peak learning rate (Lr), and total training steps (Steps). The entire framework has 545.7M parameters and requires 79.2 GMACs per second.

\rrev{\textbf{Inference details}: UniPASE operates in offline mode only. Despite containing over 500M parameters, it achieves very fast inference. On an NVIDIA RTX 4090 GPU, processing a 1-second input at 48~kHz requires approximately 44~ms with a memory footprint around 3.2~GB; for a 4-second input at 48~kHz, inference takes roughly 75~ms with about 4.6~GB of memory. We observe that both inference time and memory usage scale approximately linearly with input duration.}

\begin{table}[t]
\centering
\caption{Detailed training configurations of each module in UniPASE. \emph{DWO} denotes \emph{DeWavLM-Omni}.}
\resizebox{\linewidth}{!}{
\begin{tabular}{@{}cccccccc@{}}
\toprule
Module & Params (M) & MACs (G/s) & Fs (kHz) & Len (s) & Bs & Lr & Steps \\ \midrule
DWO & 315.44 & 18.08 & 16 & 4 & 20 & 1e-4 & 100k \\ 
Adapter & 113.73 & 5.69 & 16 & 4 & 64 & 2e-4 & 100k \\ 
Vocoder & 113.73 & 5.69 & 16 & 1 & 40 & 2e-4 & 200k \\ 
PostNet & 2.77 & 49.73 & 48 & 2 & 3 & 2e-4 & 100k \\ \bottomrule
\end{tabular}
}
\label{tab:conf}
\end{table}

\begin{table*}[t]
\caption{Comparison results on the DNS 2020 \textit{no-reverb} test set. \textit{P} and \textit{G} in the \textit{Type} column denote \textit{predictive} and \textit{generative}, respectively.}
\centering
\begin{tabular}{cccccccccc}
\toprule
Model & Type & DNSMOS $\uparrow$ & UTMOS $\uparrow$ & PESQ $\uparrow$ & ESTOI $\uparrow$ & SBS $\uparrow$ & LPS $\uparrow$ & SpkSim $\uparrow$ & dWER (\%) $\downarrow$ \\ \midrule
Noisy & - & 2.48 & 2.36 & 1.58 & 0.81 & 0.80 & 0.90 & 0.94 & 3.51 \\
Clean & - & 3.28 & 4.14 & 4.50 & 1.00 & 1.00 & 1.00 & 1.00 & 0.00 \\ \midrule
TF-GridNet & P & 3.34 & 3.86 & \textbf{3.18} & \textbf{0.94} & 0.91 & \textbf{0.97} & 0.94 & 2.86 \\
StoRM & G & 3.31 & 3.73 & 2.74 & 0.92 & 0.89 & 0.95 & 0.93 & 4.41 \\
LLaSE-G1 & G & \textbf{3.42} & 3.84 & 1.77 & 0.70 & 0.84 & 0.90 & 0.77 & 12.15 \\
AnyEnhance & G & \textbf{3.42} & 3.96 & 2.95 & 0.91 & 0.91 & 0.96 & 0.95 & 4.58 \\
PASE & G & 3.39 & 3.95 & 2.42 & 0.88 & 0.93 & \textbf{0.97} & 0.94 & 2.71 \\ \midrule
UniPASE & G & 3.40 & \rev{\textbf{4.06}} & \rev{3.05} & 0.93 & \textbf{0.94} & \rev{\textbf{0.97}} & \textbf{0.96} & \rev{\textbf{2.17}} \\ 

\bottomrule
\end{tabular}
\label{tab:dns2020_no_reverb}
\end{table*}

\begin{table*}[t]
\caption{Comparison results on the DNS 2020 \textit{with-reverb} test set. \textit{P} and \textit{G} in the \textit{Type} column denote \textit{predictive} and \textit{generative}, respectively.}
\centering
\begin{tabular}{cccccccccc}
\toprule
Model & Type & DNSMOS $\uparrow$ & UTMOS $\uparrow$ & PESQ $\uparrow$ & ESTOI $\uparrow$ & SBS $\uparrow$ & LPS $\uparrow$ & SpkSim $\uparrow$ & dWER (\%) $\downarrow$ \\ \midrule
Noisy & - & 1.39 & 1.30 & 1.16 & 0.35 & 0.61 & 0.63 & 0.70 & 10.23 \\
Clean & - & 3.28 & 4.14 & 4.50 & 1.00 & 1.00 & 1.00 & 1.00 & 0.00 \\ \midrule
TF-GridNet & P & 2.63 & 1.42 & 1.51 & 0.49 & 0.77 & 0.88 & 0.70 & 8.86 \\
StoRM & G & 2.87 & 1.84 & 1.39 & 0.49 & 0.61 & 0.60 & 0.60 & 49.65 \\
LLaSE-G1 & G & \textbf{3.35} & 2.90 & 1.20 & 0.55 & 0.71 & 0.70 & 0.55 & 41.66 \\
AnyEnhance & G & 3.20 & 2.75 & \textbf{1.79} & 0.65 & 0.80 & 0.87 & 0.70 & 14.16 \\
PASE & G & 2.75 & 1.61 & 1.41 & 0.47 & 0.81 & 0.90 & 0.60 & 9.78 \\ \midrule
UniPASE & G & \rev{3.33} & \rev{\textbf{3.62}} & \rev{1.74} & \rev{\textbf{0.76}} & \rev{\textbf{0.87}} & \rev{\textbf{0.93}} & \rev{\textbf{0.79}} & \rev{\textbf{8.16}} \\ 
\bottomrule
\end{tabular}
\label{tab:dns2020_with_reverb}
\end{table*}

\subsection{Comparison with Baselines}
\subsubsection{On the DNS 2020 test set}

The comparison results on the DNS 2020 no-reverb test set are presented in Table~\ref{tab:dns2020_no_reverb}. As expected, the predictive TF-GridNet achieves strong speaker and linguistic fidelity, obtaining a high SpkSim of 0.94 and a low dWER of 2.86\%. In contrast, several generative approaches, such as StoRM, LLaSE-G1, and AnyEnhance, exhibit noticeable linguistic degradation, with dWER values higher than those of the noisy speech, indicating a tendency toward hallucination. Notably, TF-GridNet also achieves competitive non-intrusive metrics, which can be attributed to the high-SNR condition of the no-reverb test set that favors conventional predictive models.

Both our previous PASE and the current UniPASE achieve lower dWER than the noisy speech, highlighting their low-hallucination characteristics. More importantly, UniPASE consistently outperforms other baselines across nearly all metrics, including perceptual quality (UTMOS \rev{4.06}), speaker similarity (SpkSim 0.96), and linguistic accuracy (dWER \rev{2.17\%}). UniPASE also attains the highest PESQ (\rev{3.05}) among generative models, demonstrating its ability to generate enhanced speech that is highly faithful at signal level under anechoic conditions.

Results on the with-reverb subset are presented in Table~\ref{tab:dns2020_with_reverb}. Note that the clean references and transcriptions are taken from the no-reverb test set, since the models are also expected to perform dereverberation. 
As shown, all models exhibit performance degradation compared with the no-reverb condition, reflecting the increased challenge posed by reverberation. \rev{In addition, StoRM shows a significant performance drop, likely due to the use of separately trained dereverberation and denoising checkpoints, which introduces a mismatch with its original training setup, making the results less comparable.}
\rev{For the predictive TF-GridNet, its perceptual quality degrades, but dWER remains robust. In contrast, generative baselines} LLaSE-G1 and AnyEnhance achieve higher DNSMOS and UTMOS scores, \rev{but} deliver more pronounced hallucinations under reverberant conditions, as evidenced by their high dWER values of 41.66\% and 14.16\%, respectively. 

\rev{PASE and UniPASE, in contrast, demonstrate strong robustness against linguistic hallucination, achieving low dWER values of 9.78\% and 8.16\%, respectively. Notably, UniPASE further improves perceptual quality over PASE, as reflected by DNSMOS (3.33 vs. 2.75) and UTMOS (3.62 vs. 1.61). Our additional experiments suggest that these perceptual gains largely arise from the use of dry training targets (i.e., clean speech without additioanl early reflections). Overall, UniPASE achieves strong perceptual quality while providing hallucination reduction that is competitive with predictive models.}

\begin{table*}[t]
\caption{Comparison results on the PLC 2024 validation set. \textit{P} and \textit{G} in the \textit{Type} column denote \textit{predictive} and \textit{generative}, respectively.}
\centering
\resizebox{\linewidth}{!}{
\begin{tabular}{cccccccccccc}
\toprule
Model & Type & DNSMOS $\uparrow$ & NISQA $\uparrow$ & UTMOS $\uparrow$ & PLCMOS $\uparrow$ & PESQ $\uparrow$ & ESTOI $\uparrow$ & SBS $\uparrow$ & LPS $\uparrow$ & SpkSim $\uparrow$ & WER (\%) $\downarrow$ \\ \midrule
Lossy & - & 2.82 & 2.10 & 2.48 & 2.52 & 1.94 & 0.76 & 0.82 & 0.78 & 0.61 & 18.10 \\
Clean & - & 3.36 & 4.01 & 3.76 & 4.22 & 4.50 & 1.00 & 1.00 & 1.00 & 1.00 & 8.25 \\ \midrule
TF-GridNet & P & 3.15 & 3.07 & 2.64 & 3.46 & 2.41 & 0.82 & 0.86 & 0.85 & \textbf{0.94} & 18.01 \\
LLaSE-G1 & G & 3.09 & 3.37 & 2.64 & 3.32 & 1.37 & 0.54 & 0.76 & 0.68 & 0.73 & 31.46 \\
UNIVERSE++ & G & 3.19 & - & - & - & 2.34 & 0.81 & - & 0.85 & - & 19.60 \\ \midrule
UniPASE & G & \textbf{3.39} & \rev{\textbf{4.34}} & \textbf{3.65} & \rev{\textbf{4.30}} & \rev{\textbf{2.53}} & \textbf{0.85} & \textbf{0.93} & \textbf{0.92} & \rev{\textbf{0.94}} & \rev{\textbf{13.55}} \\ \bottomrule
\end{tabular}
}
\label{tab:plc2024}
\end{table*}

\subsubsection{On the PLC 2024 validation set}
The PLC performance is summarized in Table~\ref{tab:plc2024}. Among the baselines, LLaSE-G1 exhibits severe hallucinations, resulting in a high WER of 31.46\%. UNIVERSE++, while attaining relatively strong perceptual and intelligibility performance (e.g., PESQ 2.34 and ESTOI 0.81), still suffers from limited linguistic correctness, with a WER of 19.6\%. The predictive TF-GridNet enhances perceptual quality and improves LPS over the lossy input, but yields a marginal reduction in WER, suggesting that while it can repair very short phoneme-level losses, it struggles with longer packet loss segments that disrupt linguistic content.

In contrast, UniPASE consistently outperforms all baselines across all metrics. It achieves the highest non-intrusive scores, substantially narrowing the gap to clean speech and even surpassing it in some measures. It also maintains a SpkSim of \rev{0.94}, indicating effective recovery of speaker characteristics in packet-loss regions. More importantly, UniPASE reduces WER from 18.10\% to \rev{13.55\%} and attains the highest SBS (0.93) and LPS (0.92), demonstrating its strong ability to recover longer bursts of missing content while preserving linguistic integrity. These results highlight the remarkable effectiveness of UniPASE for high-performance PLC.

\begin{table*}[t]
\caption{Comparison results on the VoiceFixer GSR test set. \textit{P} and \textit{G} in the \textit{Type} column denote \textit{predictive} and \textit{generative}, respectively.}
\centering
\begin{tabular}{ccccccccccc}
\toprule
Model & Type & DNSMOS $\uparrow$ & NISQA $\uparrow$ & UTMOS $\uparrow$ & PESQ $\uparrow$ & ESTOI $\uparrow$ & SBS $\uparrow$ & LPS $\uparrow$ & SpkSim $\uparrow$ & dWER (\%) $\downarrow$ \\ \midrule
Noisy & - & 2.39 & 2.67 & 2.63 & 1.94 & 0.58 & 0.76 & 0.84 & 0.76 & 9.50  \\
Clean & - & 3.10 & 4.28 & 3.97 & 4.50 & 1.00 & 1.00 & 1.00 & 1.00 & 0.00 \\ \midrule
TF-GridNet & P & 3.06 & 3.77 & 3.41 & \textbf{2.47} & 0.68 & 0.83 & 0.88 & 0.80 & 9.93 \\
VoiceFixer & G & 3.00 & 4.16 & 3.44 & 2.04 & 0.56 & 0.83 & 0.80 & 0.67 & 12.64 \\
AnyEnhance & G & \textbf{3.14} & 4.31 & 3.75 & 2.45 & 0.67 & 0.83 & 0.88 & 0.67 & 10.10 \\ \midrule
UniPASE & G & \rev{3.09} & \rev{\textbf{4.37}} & \rev{\textbf{3.89}} & \rev{\textbf{2.47}} & \rev{\textbf{0.69}} & \rev{\textbf{0.89}} & \rev{\textbf{0.91}} & \rev{\textbf{0.81}} & \rev{\textbf{8.21}}  \\ \bottomrule
\end{tabular}
\label{tab:voicefixer_gsr}
\end{table*}

\subsubsection{On the VoiceFixer GSR test set}
The comparison results for GSR performance are reported in Table~\ref{tab:voicefixer_gsr}. \rev{UniPASE achieves the highest scores across most perceptual metrics, including NISQA (4.37), UTMOS (3.89), and PESQ (2.47), while also attaining the lowest dWER (8.21\%) and the highest SBS (0.89) and LPS (0.91)}, demonstrating strong preservation of linguistic, phonetic, and semantic integrity. Its SpkSim score \rev{(0.81) is the highest across all baselines, indicating robust speaker fidelity. Overall, these results show that UniPASE consistently outperforms other models, achieving high perceptual quality while maintaining low hallucination, thereby ensuring reliable speech restoration.}

\begin{table*}[t]
\caption{Comparison results on the URGENT 2025 \textit{non-blind} test set. Numbers in the \textit{Rank} column indicate the final rankings in the \textit{blind} test set. \textit{P}, \textit{G}, and \textit{P+G} in the \textit{Type} column denote \textit{predictive}, \textit{generative}, and \textit{hybrid} methods, respectively.}
\centering
\resizebox{\linewidth}{!}{
\begin{tabular}{cccccccccccc}
\toprule
Model/Team & Rank & Type & DNSMOS $\uparrow$ & NISQA $\uparrow$ & UTMOS $\uparrow$ & PESQ $\uparrow$ & ESTOI $\uparrow$ & SBS $\uparrow$ & LPS $\uparrow$ & SpkSim $\uparrow$ & CER (\%) $\downarrow$ \\ \midrule
Noisy & - & - & 1.84 & 1.69 & 1.56 & 1.37 & 0.61 & 0.75 & 0.62 & 0.63 & 18.71  \\
Clean & - & - & 2.94 & 3.25 & 2.51 & 4.50 & 1.00 & 1.00 & 1.00 & 1.00 & 4.34 \\ \midrule
BSRNN-FAN  & 1 & P & 3.01 & 3.41 & 2.40 & \textbf{2.95} & \textbf{0.86} & \textbf{0.91} & \textbf{0.86} & \textbf{0.85} & \textbf{11.08} \\
TS-URGENet  & 2 & P+G & 3.00 & 3.45 & 2.31 & 2.74 & 0.84 & 0.89 & 0.84 & 0.83 & 12.06 \\
FUSE & 3 & P+G & 3.02 & 3.28 & 2.34 & 2.63 & 0.82 & 0.88 & 0.82 & 0.82 & 13.85 \\
USEMamba & 5 & P+G & 3.01 & 3.21 & 2.30 & 2.79 & 0.85 & 0.90 & 0.85 & 0.84 & 11.95 \\ \midrule
TF-GridNet & 10 & P & 2.94 & 2.89 & 2.11 & 2.43 & 0.80 & 0.86 & 0.79 & 0.80 & 15.04 \\
\texttt{wataru9871}~\cite{URGENT2025} & 13 & G & 3.18 & 4.01 & 2.78 & 1.36 & 0.56 & 0.82 & 0.73 & 0.51 & 20.30 \\ \midrule
UniPASE & - & G & \rev{\textbf{3.26}} & \rev{\textbf{4.18}} & \rev{\textbf{2.97}} & \rev{2.12} & \rev{0.70} & \rev{0.89} & 0.84 & 0.81 & \rev{12.90} \\ 
\bottomrule
\end{tabular}
}
\label{tab:urgent2025}
\end{table*}

\subsubsection{On the URGENT 2025 non-blind test set}
The comparison results on the URGENT 2025 non-blind test set are reported in Table~\ref{tab:urgent2025}. As a purely generative approach, UniPASE differs fundamentally from the top-ranking systems, most of which adopt predictive or hybrid predictive-generative strategies. Despite this, UniPASE achieves competitive performance across a wide range of evaluation metrics. In particular, it attains \rev{significantly} stronger non-intrusive perceptual scores \rev{(DNSMOS 3.26, NISQA 4.18 and UTMOS 2.97)} while maintaining competitive speaker fidelity (SpkSim 0.81) and linguistic integrity, as reflected by SBS (\rev{0.89}), LPS (0.84), and a relatively low CER (\rev{12.90\%}). Although UniPASE underperforms the leading hybrid systems in PESQ and ESTOI, this behavior is expected, as these metrics are known to be sensitive to imperceptible signal-level differences and are less suitable for generative approaches~\cite{PESQ_unreliable, LPS}.

When compared with the only other purely generative system, \texttt{wataru9871}, UniPASE exhibits clear advantages \rev{not only in perceptual quality but also} in mitigating hallucinations. \rev{Specifically, \texttt{wataru9871} attains} a high CER of 20.30\% and a low SpkSim of 0.51. In contrast, UniPASE preserves linguistic content and speaker characteristics more effectively, achieving a CER of \rev{12.90\%} and a SpkSim of 0.81, \rev{indicating robust performance with low hallucination}. 

Beyond the URGENT 2025 evaluation, UniPASE also serves as the backbone of our submission to the subsequent URGENT 2026 challenge. Specifically, we construct a simple extension by integrating UniPASE with the predictive TF-GridNet to form a hybrid framework, \rrev{which consists of a generative branch (DeWavLM-Omni, Adapter, Vocoder), a predictive TF-GridNet branch, and a PostNet for fusion and BWE. Only the PostNet is retrained to accommodate dual inputs, while all other modules remain frozen. Further details are provided in~\cite{GAP-URGENet}.} This extended system achieved 1st place in the objective evaluation in URGENT 2026\footnote{\url{https://urgent-challenge.com/competitions/15\#results}. Our team name for the competition was \texttt{WR}.}, providing further evidence of the effectiveness and extensibility of UniPASE as a generative backbone for hybrid predictive-generative designs.

\begin{table*}[t]
\caption{Ablation results on the URGENT 2025 \textit{non-blind} test set. $^\dagger$denotes \rev{train and} inference on \textit{clean} speech. 
\textit{Voc}, \textit{DWO}, \textit{Ada}, and \textit{Pos} refers to \textit{Vocoder}, \textit{DeWavLM-Omni}, \textit{Adapter}, and \textit{PostNet}, respectively.}
\centering
\resizebox{\linewidth}{!}{
\begin{tabular}{ccccccccccc}
\toprule
ID & Model & DNSMOS $\uparrow$ & NISQA $\uparrow$ & UTMOS $\uparrow$ & PESQ $\uparrow$ & ESTOI $\uparrow$ & SBS $\uparrow$ & LPS $\uparrow$ & SpkSim $\uparrow$ & CER (\%) $\downarrow$ \\ \midrule
- & Noisy & 1.84 & 1.69 & 1.56 & 1.37 & 0.61 & 0.75 & 0.62 & 0.63 & 18.71  \\
- & Clean & 2.94 & 3.25 & 2.51 & 4.50 & 1.00 & 1.00 & 1.00 & 1.00 & 4.34 \\ \midrule
1 & Voc-$\text{R}_{\text{P}}$$^\dagger$ & 3.26 & 3.93 & 2.76 & 1.29 & 0.68 & 0.92 & 0.90 & 0.62 & 5.11 \\
2 & Voc-$\text{R}_{\text{A}}$$^\dagger$ & 2.96 & 3.56 & 2.45 & 3.47 & 0.94 & 0.98 & 0.95 & 0.94 & 4.79 \\ \midrule
3 & DWO(w/o prior)+Voc-$\text{R}_{\text{P}}$ & \rev{3.29} & \rev{4.08} & \rev{2.93} & \rev{1.20} & \rev{0.48} & \rev{0.78} & 0.62 & 0.42 & \rev{34.62} \\
4 & DWO(w/o PLD)+Voc-$\text{R}_{\text{P}}$ & \rev{3.33} & \rev{4.21} & \rev{3.19} & \rev{1.27} & \rev{0.57} & \rev{0.86} & 0.79 & \rev{0.55} & \rev{16.79} \\
5 & DWO+Voc-$\text{R}_{\text{P}}$ & \rev{\textbf{3.35}} & \rev{4.25} & \rev{\textbf{3.30}} & \rev{1.27} & \rev{0.58} & \rev{0.88} & 0.83 & \rev{0.56} & \rev{\textbf{12.80}} \\ \midrule
6 & DWO+Voc-dual & \rev{3.22} & \rev{3.95} & \rev{2.64} & \rev{2.00} & \rev{0.71} & \rev{0.88} & \rev{0.83} & \rev{0.77} & \rev{13.12} \\
7 & DWO+Ada(w/o \rrev{MRRD})+Voc-$\text{R}_{\text{A}}$ & \rev{3.28} & \rev{3.60} & \rev{2.98} & \rev{\textbf{2.15}} & \rev{\textbf{0.70}} & \rev{\textbf{0.89}} & 0.84 & \rev{0.76} & \rev{12.81} \\
8 & DWO+Ada+Voc-$\text{R}_{\text{A}}$  & \rev{3.26} & \rev{\textbf{4.26}} & \rev{2.97} & \rev{2.12} & \rev{0.70} & \rev{\textbf{0.89}} & \textbf{0.84} & 0.80 & \rev{12.87} \\ \midrule
9 & DWO+Ada+Voc-$\text{R}_{\text{A}}$+Pos  & \rev{3.26} & \rev{4.18} & \rev{2.97} & \rev{2.12} & \rev{0.70} & \rev{\textbf{0.89}} & \textbf{0.84} & \textbf{0.81} & \rev{12.90} \\ 
\bottomrule
\end{tabular}
}
\label{tab:ablation}
\end{table*}

\section{\rev{Ablation and Analysis}}
\subsection{Effects of Key Design Choices}
In this section, we systematically explore the effects of our key designs, including: (1) the Vocoder based on acoustic representations, (2) DeWavLM-Omni with the inherited phonological prior and the auxiliary PLD algorithm, (3) the Adapter together with the \rrev{MRRD}, and (4) the PostNet. \rrev{The ablation study is primarily conducted on the URGENT 2025 non-blind test set (Table~\ref{tab:ablation}), which offers the most comprehensive evaluation scenario. We further note that key ablations were also performed on the VoiceFixer GSR dataset, where the observed trends are consistent, providing additional support for the effectiveness of these components.}

\subsubsection{Vocoder}
We train two Vocoder variants based on either clean phonetic representations ($\text{R}_{\text{P}}$) or acoustic representations ($\text{R}_{\text{A}}$) to validate that $\text{R}_{\text{A}}$ contains rich details necessary for high-fidelity waveform reconstruction. The Vocoder based on $\text{R}_{\text{P}}$ (ID~1) achieves superior non-intrusive metrics and competitive linguistic integrity, but suffers in PESQ, ESTOI, and speaker fidelity, indicating its unsuitability for waveform synthesis. In contrast, the Vocoder based on $\text{R}_{\text{A}}$ (ID~2) delivers superior performance across all metrics, notably achieving PESQ 3.47, SpkSim 0.94, and CER 4.79\%, demonstrating the effectiveness of using $\text{R}_{\text{A}}$ for waveform reconstruction.

It should be noted that ID~1 uses the pretrained WavLM $\text{R}_{\text{P}}$, while ID~2 uses the DeWavLM-Omni $\text{R}_{\text{A}}$. This setup provides each subsequent module with a vocoder tailored for waveform reconstruction, due to the fact that DeWavLM-Omni aims to produce clean WavLM $\text{R}_{\text{P}}$, whereas the Adapter targets clean DeWavLM-Omni $\text{R}_{\text{A}}$. Consequently, it allows the isolated evaluation of each module's performance. \rev{Regarding the Adapter training target, our preliminary experiments showed that using pretrained WavLM $\text{R}_{\text{A}}$ or DeWavLM-Omni $\text{R}_{\text{A}}$ yields comparable performance; we adopt the latter for improved training efficiency and framework consistency.}

\subsubsection{DeWavLM-Omni} 
To assess the contribution of the phonological prior, we compare two DeWavLM-Omni variants: one initialized from the pre-trained WavLM to inherit the prior (ID~5), and the other trained from scratch without this prior (ID~3). It can be seen that incorporating the phonological prior leads to substantial improvements in linguistic integrity, as evidenced by markedly lower CER (\rev{12.80\%} vs. \rev{34.62\%}) and higher LPS (0.83 vs. 0.62) and SBS (\rev{0.88} vs. \rev{0.78}). These pronounced performance gaps underscore the critical role of the prior in mitigating linguistic hallucinations.

To further examine the effect of the PLD algorithm in DeWavLM-Omni, we evaluate a variant \rev{in which the PLD detector is removed during both training and inference} (ID~4). Compared with ID~4, incorporating PLD (ID~5) leads to consistent improvements in UTMOS (\rev{3.30} vs. \rev{3.19}), SBS (\rev{0.88} vs. \rev{0.86}), LPS (0.83 vs. 0.79), and CER (\rev{12.80\%} vs. \rev{16.79\%}). 
Our manual inspection reveals that while DeWavLM-Omni can largely recover short bursts of packet loss even without PLD, it occasionally fails on long bursts. We also observe that removing PLD leads to noticeable degradations even in utterances without packet loss but affected by other distortions, such as noise or reverberation.
This suggests that the explicit masking mechanism in PLD during training promotes contextual modeling by forcing the network to infer missing content from surrounding cues, thereby reinforcing the phonological prior learned during WavLM pre-training.

\subsubsection{Adapter}
To validate the effect of the Adapter, we compare two strategies: (1) direct waveform reconstruction from dual-stream representations using a jointly trained vocoder, as in PASE (ID~6), and (2) reconstruction from Adapter-enhanced~$\text{R}_{\text{A}}$ using the pre-trained $\text{R}_{\text{A}}$-based Vocoder (ID~7). For a fair comparison, the vocoder in ID~6 is scaled to match the total parameter count and computational cost of the combined Adapter and Vocoder.
As shown, introducing the Adapter leads to consistent improvements in perceptual quality (UTMOS \rev{2.98} vs. \rev{2.64}, PESQ \rev{2.15} vs. \rev{2.00}), demonstrating its effectiveness in refining acoustic representations for downstream waveform reconstruction.

When incorporating the \rrev{MRRD} discriminator (ID~8), NISQA improves markedly from \rev{3.60} to \rev{4.26}, accompanied by a steady increase in SpkSim from \rev{0.76} to 0.80, \rev{indicating improved perceptual quality and fidelity. In contrast, DNSMOS and UTMOS show only minor changes, suggesting certain perceptual improvements may not be equally captured by different objective metrics. Given this inconsistency, we further conduct subjective evaluations to assess perceptual quality, as presented in Section~\ref{sec:subjective}.}

\subsubsection{PostNet}
The effects of incorporating PostNet are reported in ID~9. Objective metrics operating at 16~kHz remain largely unchanged, as the low-frequency bands (up to 8~kHz bandwidth) are directly copied. In contrast, the 48-kHz NISQA exhibits a slight decrease, likely due to imperfect coherence between the generated high-frequency components and the original low-frequency bands. \rev{To complement this, we further conduct subjective evaluations, as detailed in Section~\ref{sec:subjective}.}

\begin{table}[t]
\caption{\rev{Pairwise CCR results on the URGENT 2025 \emph{non-blind} test set. Results are reported as mean $\pm$ 95\% confidence intervals (CI).}}
\centering
\begin{tabular}{cccc}
\toprule
\rev{Pair} & \rev{ID} & \rev{Model} & \rev{CMOS $\uparrow$} \\
\midrule
\multirow{2}{*}{\rrev{MRRD}}
& \rev{7} & \rev{DWO+Ada(w/o \rrev{MRRD})+Voc-$\text{R}_{\text{A}}$} & \rev{0.00 $\pm$ 0.00} \\
& \rev{8} & \rev{DWO+Ada+Voc-$\text{R}_{\text{A}}$} & \rev{+1.23 $\pm$ 0.10} \\
\midrule
\multirow{2}{*}{\rev{PostNet}}
& \rev{8} & \rev{DWO+Ada+Voc-$\text{R}_{\text{A}}$} & \rev{0.00 $\pm$ 0.00} \\
& \rev{9} & \rev{DWO+Ada+Voc-$\text{R}_{\text{A}}$+Pos} & \rev{+1.65 $\pm$ 0.11} \\
\bottomrule
\end{tabular}
\label{tab:ablation_cmos}
\end{table}

\subsubsection{\rev{Subjective Evaluations}}
\label{sec:subjective}
\rev{As the objective metrics do not consistently reflect the improvements brought by the \rrev{MRRD} discriminator and PostNet modules, we further conduct subjective evaluations to assess their perceptual impact. We adopt the comparison category rating (CCR) method defined in ITU-T P.800~\cite{ITU-P.808}, which is well-suited for capturing subtle perceptual differences~\cite{CCR_Crowdsourcing}. In each trial, listeners were presented with a pair of stimuli derived from the same utterance, corresponding to systems with and without the target module. Listeners were asked to rate the perceptual quality of the second stimulus relative to the first on a 7-point scale ranging from -3 (much worse) to +3 (much better). Here, perceptual quality refers to the overall listening experience, including naturalness, clarity, and high-frequency content. The average score is reported as the comparative mean opinion score (CMOS). The evaluation was conducted on 40 audio pairs with 24 listeners.}

\rev{The results are summarized in Table~\ref{tab:ablation_cmos}. The \rrev{MRRD} module achieves a CMOS of +1.23, indicating a substantial improvement over the baseline. The PostNet module yields a CMOS of +1.65, demonstrating improved high-frequency reconstruction and overall perceptual quality. Both improvements are statistically significant (p $\le$ 0.05). Representative audio examples are visualized in Appendix~\ref{app:demo} for intuitive illustration.}

\subsection{Robustness of PLC Performance}
In this section, we systematically analyze the PLC performance of UniPASE on the PLC 2024 validation set, which covers a wide range of packet-loss conditions along two factors: (1) \textit{loss fraction}, the proportion of lost packets in an utterance, and (2) \textit{longest burst length}, the maximum number of consecutive lost packets in an utterance, hereafter occasionally referred to as \textit{burst} for brevity. The set spans multiple loss-fraction levels and burst ranges, from short and medium gaps to extremely long losses exceeding one second, enabling an evaluation of system robustness across diverse scenarios. To enable a concise yet comprehensive evaluation, we focus on four key metrics: PLCMOS for perceptual quality, SpkSim for speaker fidelity, LPS for phoneme integrity, and WER for content correctness. 

\subsubsection{Impact of Loss Fraction}
Table~\ref{tab:analysis_plc_fraction} shows the PLC performance of UniPASE across different packet-loss fractions, with each cell reporting the lossy baseline \rev{(i.e., unprocessed audio) on the left and the UniPASE output on the right}. As shown, UniPASE consistently outperforms the lossy baseline across all fractions. Notably, its PLCMOS remains high and stable across different fraction ranges, suggesting that perceptual quality is relatively easy to recover. In contrast, SpkSim, LPS, and WER remain robust at fractions up to 40\% but degrade noticeably under extreme conditions ($>$40\% fraction).

\begin{table}[t]
\caption{PLC performance across diverse loss fractions, with each cell showing two values: left for the lossy baseline, right for UniPASE.}
\centering
\begin{tabular}{ccccc}
\toprule
Fraction & PLCMOS $\uparrow$ & SpkSim $\uparrow$ & LPS $\uparrow$ & WER (\%) $\downarrow$ \\ \midrule
0-10\% & 3.72 / \rev{\textbf{4.33}} & 0.95 / \rev{\textbf{0.96}} & \textbf{0.96} / \textbf{0.96} & 11.26 / \rev{\textbf{11.10}} \\
10-20\% & 2.81 / \rev{\textbf{4.29}} & 0.78 / \rev{\textbf{0.95}} & 0.90 / \textbf{0.93} & 14.74 / \rev{\textbf{12.58}} \\
20-30\% & 2.08 / \rev{\textbf{4.28}} & 0.61 / \rev{\textbf{0.93}} & 0.82 / \textbf{0.91} & 18.10 / \rev{\textbf{13.64}} \\
30-40\% & 1.70 / \rev{\textbf{4.28}} & 0.36 / \rev{\textbf{0.94}} & 0.72 / \textbf{0.90} & 18.34 / \rev{\textbf{14.24}} \\
40-100\% & 1.36 / \rev{\textbf{4.29}} & 0.09 / \rev{\textbf{0.88}} & 0.40 / \rev{\textbf{0.84}} & 34.62 / \rev{\textbf{21.09}} \\
\bottomrule
\end{tabular}
\label{tab:analysis_plc_fraction}
\end{table}

\begin{table}[t]
\caption{PLC performance across different longest burst lengths, with each cell showing two values: left for the lossy baseline, right for UniPASE.}
\centering
\begin{tabular}{ccccc}
\toprule
Burst & PLCMOS $\uparrow$ & SpkSim $\uparrow$ & LPS $\uparrow$ & WER (\%) $\downarrow$ \\ \midrule
0-6 & 2.99 / \rev{\textbf{4.35}} & 0.85 / \rev{\textbf{0.95}} & 0.91 / \textbf{0.97} & 9.32 / \rev{\textbf{9.28}} \\
6-25 & 2.15 / \rev{\textbf{4.33}} & 0.54 / \rev{\textbf{0.93}} & 0.76 / \textbf{0.94} & 16.17 / \rev{\textbf{11.11}} \\
25-50 & 2.36 / \rev{\textbf{4.26}} & 0.48 / \rev{\textbf{0.93}} & 0.72 / \rev{\textbf{0.87}} & 27.31 / \rev{\textbf{18.58}} \\
50-150 & 2.89 / \rev{\textbf{4.13}} & 0.48 / \rev{\textbf{0.91}} & 0.70 / \rev{\textbf{0.76}} & 32.81 / \rev{\textbf{28.31}} \\
\bottomrule
\end{tabular}
\label{tab:analysis_plc_burst}
\end{table}

\subsubsection{Impact of Longest Burst Length}
Table~\ref{tab:analysis_plc_burst} shows PLC performance across different longest burst lengths. PLCMOS remains high and stable across bursts, SpkSim is robust for short to medium bursts (0–50 packets), while LPS and WER remain stable only up to 25 packets, degrading gradually for 25–50 packets and substantially for bursts exceeding 50 packets. Notably, although the model was trained with bursts of up to 10 lost packets, it generalizes well to bursts of up to 25 packets, demonstrating strong generalization capability.

\subsubsection{Comparative Analysis of Loss Fraction and Burst Length}
A combined comparison highlights that loss fraction has a stronger immediate impact on audio quality than burst length. For the lossy baselines, extreme packet-loss fractions (40–100\%) primarily disrupt phoneme-level information, with LPS dropping to 0.40, PLCMOS to 1.36, and SpkSim to 0.09. The phoneme-level degradation also results in a high overall WER of 34.62\%. In contrast, long-burst scenarios mainly impair content correctness, with WER rising to 32.81\%, while PLCMOS, SpkSim, and LPS remain relatively less affected.

Interestingly, despite their stronger initial impact, high-fraction losses are generally easier to restore. For example, even for 40–100\% loss fractions, UniPASE maintains high LPS (\rev{0.84}), PLCMOS (\rev{4.29}), SpkSim (\rev{0.88}), and a relatively low WER (\rev{21.09\%}). Long bursts, by contrast, present a greater challenge: for bursts of 50–150 packets, LPS drops to \rev{0.76}, and WER rises to \rev{29.31\%}, reflecting that consecutive missing packets limit the available context for accurate reconstruction, making content-level recovery more difficult. 

\subsubsection{Robustness to Linguistic Hallucination}
\begin{figure}
    \centering
    \includegraphics[width=0.9\linewidth]{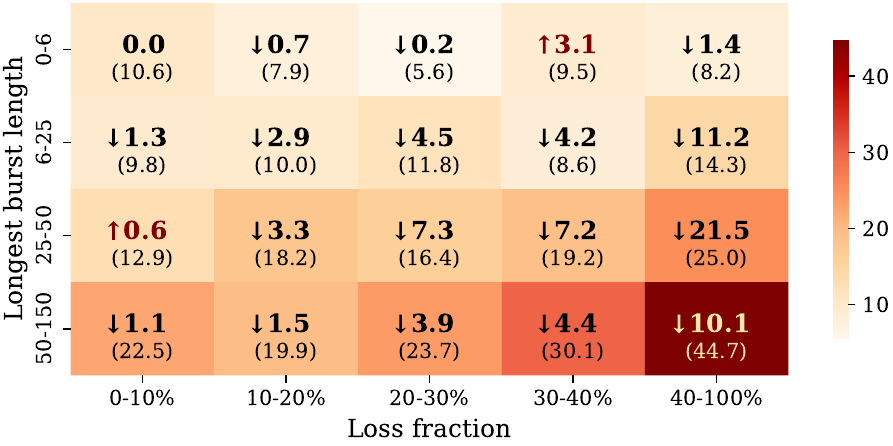}
    \caption{WER scores across different loss fractions and longest burst lengths. \rev{In each cell, the top value shows the WER change relative to the lossy baseline, and the bottom value shows the absolute WER of the UniPASE output.}}
    \label{fig:analysis_plc_heatmap}
\end{figure}

We further assess the applicability and robustness of UniPASE by examining its most critical aspect---linguistic hallucination, measured by WER---across varying loss fractions and burst lengths. \rev{The results are presented in Fig.~\ref{fig:analysis_plc_heatmap}, where each cell reports the change in WER of the UniPASE output relative to the lossy baseline, with the value in parentheses indicating the absolute WER of the UniPASE output}. UniPASE demonstrates strong robustness in typical scenarios. For short bursts (0-6 packets), \rev{WER changes remain near zero across all fractions, largely because the ASR model is robust to such short losses. For} moderate bursts (6-25 packets), \rev{the reduction in WER improves from 1.3 to 11.2 points as the loss fraction increases, maintaining the final WER at a relatively low level ($\le$ 14.3\%), indicating the effectiveness of PLC. For longer bursts (25-50 packets), the reduction in WER continues to increase, reaching 21.5 points at the most extreme loss fraction (40-100\%), highlighting that PLC remains robust even under these challenging conditions}. Very long bursts (50-150 packets) present the greatest challenge, \rev{where the WER reduction begins to degrade and the absolute WER increases. The highest WER reaches 44.7\%, a level generally considered unusable, despite a noticeable reduction of 10.1 points over the lossy baseline}. \rev{Overall,} these results confirm that UniPASE \rev{preserves} content integrity and remains robust under most realistic conditions, i.e., bursts of $\le$25 packets and loss fractions $\le$40\%.

\subsection{Cross-Language Generalizability of UniPASE}
In this section, we analyze the cross-language generalizability of UniPASE, \rev{as} its core module, DeWavLM-Omni, is built upon WavLM pre-trained exclusively on English data. \rev{Rather than focusing on generalization to completely unseen languages, we examine whether DeWavLM-Omni remains effective for languages beyond English.} Experiments are conducted on the URGENT 2025 non-blind test set, which includes five languages: 100 Chinese, 300 English, 200 French, 200 German, and 200 Spanish utterances. Our analysis spans three aspects across languages: (1) information captured in the acoustic representations, (2) information encoded in the phonetic representations, and (3) the phonological prior. PESQ and SpkSim are used as proxies to evaluate whether low-level acoustic details and speaker-related characteristics are preserved across languages. LPS and $\Delta$CER (the difference in CER between the \rev{evaluated} and clean speech) are used as proxies to evaluate whether phoneme- and character-level information generalizes across languages. \rev{The use of $\Delta$CER helps mitigate potential bias introduced by the ASR model across different languages.}

\subsubsection{Acoustic representation}
\begin{figure}
    \centering
    \subfigure[]{
        \includegraphics[width=0.95\linewidth]{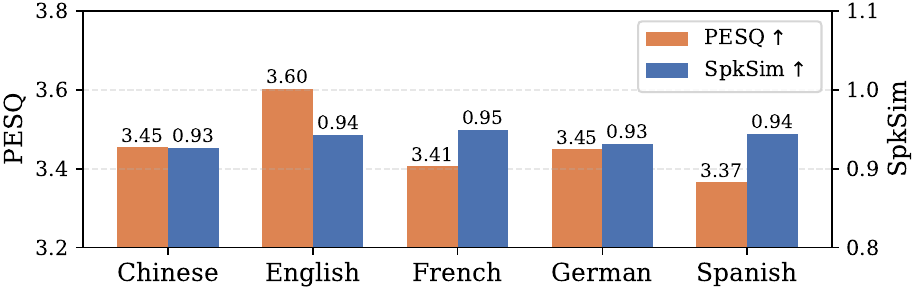}
    }
    \subfigure[]{
        \includegraphics[width=0.95\linewidth]{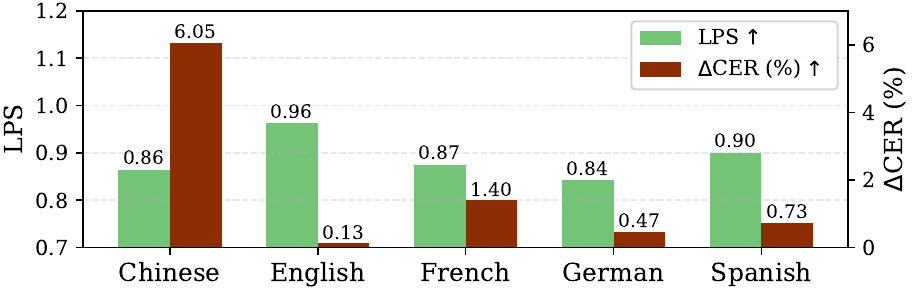}
    }
    \subfigure[]{
        \includegraphics[width=0.95\linewidth]{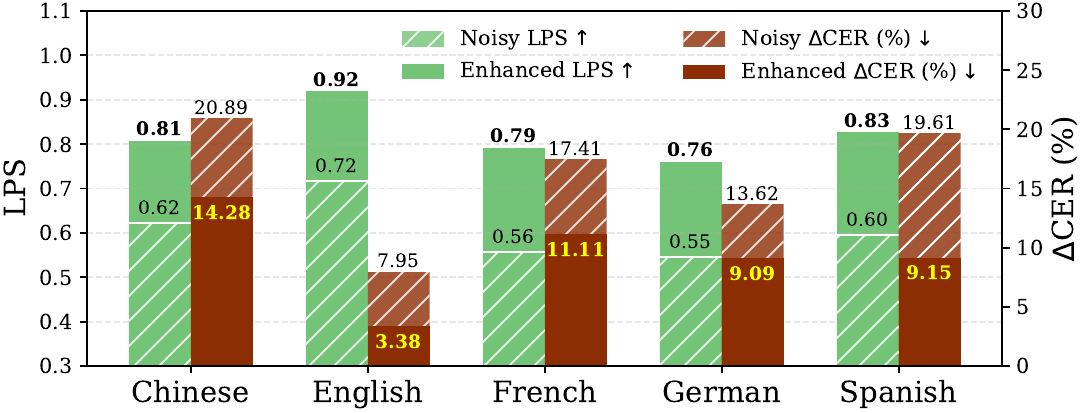}
    }

    \caption{Cross-language analysis of (a) acoustic representations, (b) phonetic representations, and (c) the phonological prior.}
    \label{fig:analysis_lang}
\end{figure}

To assess the cross-language generalizability of acoustic information captured in the acoustic representations, we adopt the same vocoder-based reconstruction setup introduced in the ablation study (ID~2 in Table~\ref{tab:ablation}) as a probe. 
As illustrated in Fig.~\ref{fig:analysis_lang}~(a), PESQ scores remain high across all languages (3.37–3.60), indicating that low-level acoustic details are well preserved. SpkSim also stays consistently strong (0.93–0.95), demonstrating that speaker characteristics are reliably retained. These results suggest that the acoustic representations carry rich low-level information that generalizes effectively across languages.

\subsubsection{Phonetic representation}
To examine the cross-language robustness of phonetic information encoded in the phonetic representations, we adopt the same vocoder-based reconstruction setup introduced in the ablation study (ID~1 in Table~\ref{tab:ablation}) as a probe. 
As shown in Fig.~\ref{fig:analysis_lang}~(b), English achieves the highest reconstruction quality, with LPS of 0.96 and a minimal $\Delta$CER of 0.13\%, reflecting near-perfect phoneme- and character-level preservation. Other languages show noticeable degradations: LPS ranges from 0.84 (German) to 0.90 (Spanish), and $\Delta$CER spans 0.47\% (German) to 6.05\% (Chinese). Despite these declines, the phonetic representations still retain a substantial amount of phonetic information, indicating that they generalize reasonably well to non-English languages, even though they were trained exclusively on English.

\subsubsection{The phonological prior}
To evaluate the cross-language transferability of the phonological prior, we analyze the enhanced speech generated by DeWavLM-Omni cascaded with a vocoder (ID~5 in Table~\ref{tab:ablation}). The results are shown in Fig.~\ref{fig:analysis_lang}~(c), where hatched bars denote the noisy input and solid bars correspond to the enhanced output. The overlap between the two bars at each language visually reflects the performance gains introduced by the enhancement.

In terms of absolute performance, English naturally achieves the best results, with an LPS score of 0.92 and a $\Delta$CER of \rev{3.38\%}. Other languages exhibit lower absolute performance: LPS ranges from 0.76 (German) to \rev{0.83} (Spanish), and $\Delta$CER from \rev{9.09\%} (\rev{German}) to \rev{14.28\%} (\rev{Chinese}). 
However, it is important to note that lower absolute performance for non-English languages does not necessarily indicate a deficiency of the model. Two factors likely contribute: (1) the noisy inputs for non-English languages are inherently more distorted in the test set, as reflected by their higher $\Delta$CER and lower LPS; (2) the CER metric, computed via a multilingual ASR, may exhibit bias towards English, making absolute scores less comparable across languages. 

When considering the relative improvements from noisy to enhanced speech, the gains brought by the phonological prior remain largely consistent across languages. Specifically, LPS increases by roughly 0.2 for all languages, indicating similar improvements in phoneme-level reconstruction. For $\Delta$CER, Spanish benefits the most, with reduction of approximately 10\%, while the other languages show consistent reductions of around \rev{6\%}. These results suggest that the phonological prior learned from English data can generalize effectively to other languages, likely because WavLM captures fundamental phoneme-level representations and knowledge shared across languages, rather than language-specific ones.

\section{Conclusions}
In this study, we present UniPASE, a universal speech enhancement framework capable of handling diverse distortions across multiple sampling rates. \rev{UniPASE employs DeWavLM-Omni to generate enhanced phonetic representations, which condition an Adapter that refines acoustic representations, followed by a vocoder for waveform reconstruction and a PostNet for flexible sampling-rate I/O, enabling high-fidelity enhancement with minimal hallucination.}
Experimental results on several evaluation datasets spanning sub-tasks and full tasks demonstrate that UniPASE achieves superior or competitive performance against SOTA approaches, highlighting its high-fidelity and low-hallucination characteristics. Further analysis confirms that UniPASE is robust and reliable across sub-tasks, effectively handling severe distortions within each task and maintaining consistent performance across multiple languages.

\bibliographystyle{IEEEtran}
\bibliography{ref.bib}

\appendix
\subsection{The Packet Loss Detection Algorithm}
\label{app:alg}

As shown in Algorithm~\ref{alg:pld}, the Packet Loss Detection (PLD) algorithm segments the input waveform into short, non-overlapping packets and identifies nearly silent ones. Specifically, for each packet, we compute the fraction of samples with amplitudes below a small threshold; if this fraction exceeds a predefined ratio, the packet is flagged as lost. This process yields a binary mask indicating the locations of missing packets, which is then used by DeWavLM-Omni for masked prediction. Notably, in our setup, the packets per second (PPS) match the frames per second (FPS) of the WavLM embeddings, allowing the PLD mask to be applied directly without any interpolation.

\begin{algorithm}[t]
\small
\caption{Packet Loss Detection (PLD)}
\label{alg:pld}
\begin{algorithmic}[1]
\Require Audio $\mathbf{x} \in \mathbb{R}^{L}$, sampling rate $f_\text{s}$, packet duration $t_\text{packet}$, amplitude threshold $\epsilon$, minimum zero ratio $r_\text{min}$
\Ensure Binary mask $\mathbf{M}_{\text{T}}$ indicating lost packet indices

\State $P \gets f_\text{s} \cdot t_\text{packet}$ \Comment{packet length in samples}
\State $N \gets \lfloor L / P \rfloor$ \Comment{number of packets}
\State Initialize mask: $\mathbf{M}_\text{T} \gets \mathbf{0}^{N}$

\For{$i = 1$ to $N$}
    \State $\mathbf{x}_i \gets \mathbf{x}[(i-1)P+1 : iP]$ \Comment{samples in packet $i$}
    \State $r_i \gets \frac{1}{P} \sum_{j=1}^{P} \mathbf{1}\{|\mathbf{x}_{i,j}| < \epsilon\}$ \Comment{compute silence ratio}
    \If{$r_i \ge r_\text{min}$}
        \State $\mathbf{M}_\text{T}[i] \gets 1$ \Comment{flag packet as lost}
    \EndIf
\EndFor

\State \Return $\mathbf{M}_{\text{T}}$
\end{algorithmic}
\end{algorithm}

\subsection{Audio Example Visualizations}
\label{app:demo}
Given that the improvements brought by the proposed \rrev{MRRD} and PostNet components are not fully captured by most objective metrics, we provide qualitative audio examples to demonstrate their effectiveness.

\begin{figure}[t]
    \centering

    \subfigure[Without \rrev{MRRD}]{
        \includegraphics[width=0.46\linewidth]{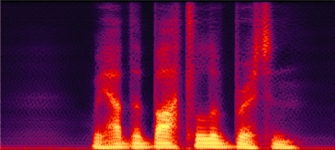}
    }
    \subfigure[With \rrev{MRRD}]{
        \includegraphics[width=0.46\linewidth]{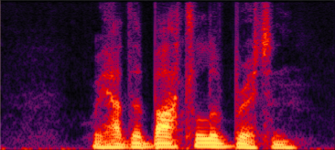}
    }
    
    \subfigure[Without PostNet]{
        \includegraphics[width=0.46\linewidth]{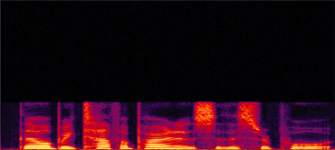}
    }
    \subfigure[With PostNet]{
        \includegraphics[width=0.46\linewidth]{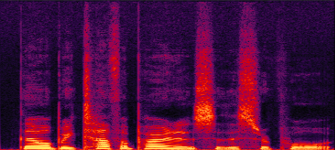}
    }
    \caption{Qualitative comparisons. Top: effect of \rrev{MRRD}. Bottom: effect of PostNet. Each row should be compared horizontally.}
    \label{fig:qualitative_samples}
\end{figure}

As demonstrated in Fig.~\ref{fig:qualitative_samples} (a) and (b), the absence of \rrev{MRRD} results in an over-smoothed spectrogram with reduced spectral variation, whereas incorporating \rrev{MRRD} produces richer spectral patterns with more visible details.
In addition, Fig.~\ref{fig:qualitative_samples} (c) and (d) show that the PostNet restores high-frequency components, demonstrating its effectiveness for BWE.

\end{document}